\newif\ifmulticol	\multicoltrue
\newif\ifshowgit	\showgittrue		
\newif\ifgitlocal	\gitlocaltrue		
\newif\ifbiblatex	\biblatexfalse		
\newif\ifbibnum		\bibnumtrue 		
\newif\iflineno		\linenofalse
\newif\iftoc		\toctrue

\newif\iflucida		\lucidafalse
\newif\ifcm			\cmfalse
\newif\iflibertine	\libertinefalse
\newif\ifcharter	\chartertrue

\newcommand*{\secnumdepth}{0}			
\newcommand*{\tocdepth}{2}				
\newcommand*{\reftitle}{References}		
\newcommand*{\mytitleskip}{-0.2in}		

\newcommand*{\mydocfontsize}{\ifcharter11pt\else\iflibertine11pt\else10pt\fi\fi}

\documentclass[\mydocfontsize]{article}

\newcommand*{\pdfauthor}{Steven A.~Frank}
\newcommand*{\pdftitle}{Receptor uptake arrays for vitamin B12, siderophores and glycans shape bacterial communities}

\input rpoly.sty

\newcommand*{\bt}{B$_{12}$}
\newcommand*{\btb}{B$_{\mathbf{12}}$}
\newcommand*{\Bo}{\textit{Bo}}
\newcommand*{\Bv}{\textit{Bv}}

\newcommand\process[1]{\medskip\noindent\textit{#1.}\space}

\newcommand*{\Ga}{\alpha}

\newcommand*{\Gs}{\sigma}

\newcount\BoxNum \BoxNum 1\relax
\makeatletter
\newcommand*{\boxlabel}[1]{%
  \protected@write \@auxout {}{\string \newlabel {box:#1}{{\the\BoxNum}}{}}%
  \advance\BoxNum 1\relax}
\makeatother

\newcommand*{\mytitle}{Receptor uptake arrays for vitamin B$_{\hbox{\sffamily\normalsize 12}}$, siderophores and glycans shape bacterial communities}

\newcommand*{\myauthor}{Steven A.\ Frank}
\newcommand*{\myaddress}{Department of Ecology and Evolutionary Biology, University of California, Irvine, CA 92697--2525, USA}
\newcommand*{\mycontact}{\href{https://stevefrank.org}{https://stevefrank.org}}

\newcommand*{\mykeywords}{Corrinoids, public goods, microbiome, systems biology, deep learning, control theory, phenotypic plasticity}

\setabstract{-0.07}{\iftoc-2.0\else1.22\fi}{%
Molecular variants of vitamin \bt, siderophores and glycans occur. To take up variant forms, bacteria may express an array of receptors. The gut microbe \textit{Bacteroides thetaiotaomicron} has three different receptors to take up variants of vitamin \bt\ and 88 receptors to take up various glycans. The design of receptor arrays reflects key processes that shape cellular evolution. Competition may focus each species on a subset of the available nutrient diversity. Some gut bacteria can take up only a narrow range of carbohydrates, whereas species such as \textit{B.~thetaiotaomicron} can digest many different complex glycans. Comparison of different nutrients, habitats, and genomes provide opportunity to test hypotheses about the breadth of receptor arrays. Another important process concerns fluctuations in nutrient availability. Such fluctuations enhance the value of cellular sensors, which gain information about environmental availability and adjust receptor deployment. Bacteria often adjust receptor expression in response to fluctuations of particular carbohydrate food sources. Some species may adjust expression of uptake receptors for specific siderophores. How do cells use sensor information to control the response to fluctuations? That question about regulatory wiring relates to problems that arise in control theory and artificial intelligence. Control theory clarifies how to analyze environmental fluctuations in relation to the design of sensors and response systems. Recent advances in deep learning studies of artificial intelligence focus on the architecture of regulatory wiring and the ways in which complex control networks represent and classify environmental states. I emphasize the similar design problems that arise in cellular evolution, control theory, and artificial intelligence. I connect those broad conceptual aspects to many testable hypotheses for bacterial uptake of vitamin \bt, siderophores and glycans.
}

\begin{document}

\mymaketitle

\ifmulticol\begin{multicols}{2}\fi

\iftoc\mytoc{-24pt}{\newpage}\fi

\section{Introduction}

Molecular variants of vitamin \bt\ occur. A bacterial cell may take up different \bt\ forms by expressing multiple receptors. Bacterial cells also express multiple receptors to take up polymorphic iron-scavenging siderophores and energy containing glycans. I emphasize broad questions about the characteristics of receptor uptake arrays for nutrient acquisition. 

How does natural selection set the number of receptor variants? How does selection tune the binding affinities of different receptor variants in an array? How does the design of receptor arrays shape the competitive and cooperative processes that define bacterial communities?

Vitamin and metabolite receptor arrays provide an exceptional model to study conditional response to the environment. Nutrient availability fluctuates. When cells can produce many different receptors, it may pay to alter receptor expression levels according to the availability of matching nutrients. 

Plastic response requires sensor arrays to perceive availability. Environmental perception must then be transduced through a regulatory network that integrates information and alters deployment of the receptor array. How do aspects of environmental fluctuation shape the evolutionary design of sensory perception and the regulatory control to achieve a conditional response? 

I discuss how perception, classification, and response through a network relate to recent breakthroughs in artificial intelligence, neural networks, and deep learning. Bacterial systems provide great opportunity to link conceptual aspects of evolutionary design and deep learning to hypotheses that can be tested by comparative genomics and by experimental laboratory studies.

I synthesize aspects of receptor arrays for vitamin \bt\ analogs, for siderophores, and for glycans. By considering these different cases together, deeper principles of receptor variety and specificity emerge. Here, I use the word `receptor' to include the various binding and transport processes that influence specificity and rate of uptake.

For uptake of vitamin \bt\ analogs, \textit{Bacteroides thetaiotaomicron} expresses three homologous variants of the associated receptor \autocite{degnan14human}. Competition experiments demonstrate that particular receptors lead to competitive dominance in the presence of particular \bt\ variants, whereas other receptors cause dominance in the presence of other \bt\ variants. Receptors apparently differ in their uptake efficacies for a variety of \bt\ analogs. \textcite{degnan14human} estimate that the human microbiome contains more than 30 receptor families for uptake of \bt-like corrinoid variants. Genomic analyses predict that a significant fraction of all bacteria and archaea require a corrinoid variant for growth, yet only a minority can produce corrinoids \autocite{rodionov03comparative,zhang09comparative}. Competition over corrinoid variants shapes the receptor arrays of individual species and the dynamics of bacterial communities. 

Siderophores are secreted molecules that bind free iron. Bacteria use siderophore receptors to take up the siderophore-iron complexes. Iron often sets a limiting resource for microbial growth. Thus, competitive and cooperative processes over siderophore uptake shape bacterial community dynamics \autocite{west07the-social,chakraborty13iron,niehus17the-evolution}. Individual bacteria may secrete more than one type of siderophore, or none at all. Bacteria typically have uptake receptors tuned for their own secreted types. In addition, bacteria often express an array of siderophore uptake receptors for types produced by other species \autocite{loper99utilization}. Bacteria, yeast, and other fungi may take up each other's secreted siderophores. The ubiquitous battle for free iron sets the design of siderophore uptake receptor arrays.

Glycans are complex carbohydrates with diverse molecular structures. Bacteria use specific receptors and digestive enzymes to catabolize particular glycans. In habitats with diverse glycan sources, such as the mammalian gut, bacteria often express broad glycan receptor arrays. For example, many species of Bacteroidetes have diverse Polysaccharide Utilization Loci (PUL) gene clusters \autocite{martens09complex,flint12microbial,grondin17polysaccharide}. Each cluster typically encodes cell surface glycan-binding proteins that capture specific glycans to initiate catabolism. Among \textit{B.~thetaiotaomicron}, \textit{B.~ovatus} and \textit{B.~cellulosilyticus} WH2, each has approximately 100 PULs. Species pairs vary in their number of shared PULs, probably reflecting the different habitats and the different competitive tunings of their receptor arrays. 

\section{Vitamin \btb\ variants}

I begin with the variety of \bt\ molecules. How are different receptors tuned to compete for taking up the diversity of \bt\ variants?

\subsection{Background} 

\bt\ variants comprise a family of corrinoid molecules \autocite{roth96cobalamin,matthews09cobalamin-and}. The cobalt-containing corrin ring defines the group and plays a key role in corrinoid coenzyme activity. I use `\bt\ variant' and `corrinoid' interchangeably. 

Genomic studies estimate that 76\% of bacterial species require a \bt\ variant. Only one-half of those species that require a \bt\ variant can synthesize their own. The other species must take up the vitamin externally. Nearly all species that require a \bt\ variant, including those that can synthesize their own, have genes that encode uptake \autocite{zhang09comparative}. 

Corrinoid synthesis is complex and energetically costly, whereas uptake is relatively inexpensive \autocite{roth96cobalamin}.  The cost difference between synthesis and uptake probably explains why uptake is so common, even among those species that can make their own.

Only prokaryotes can make corrinoids. Free corrinoids most likely originate from release of intracellular components following prokaryotic cell death. No active secretion is known. Corrinoids may also cycle among various prokaryotic and eukaryotic consumers and the environment.

\subsection{Corrinoids, growth and competition} 

Many studies implicate corrinoids as a limiting growth factor for prokaryotes and marine eukaryotic plankton \autocite{zhang09comparative,sanudo-wilhelmy14the-role}. Most examples are based on three lines of evidence: absence of corrinoid-producing genes, presence of essential corrinoid-requiring genes without alternative corrinoid-independent pathways, and very low levels of free corrinoids. Additionally, supplementation of phytoplankton communities with \bt\ variants sometimes stimulates growth or significantly alters community composition \autocite{koch11the-effect,sanudo-wilhelmy14the-role}. 

However, it can be difficult to interpret the importance of \bt\ variants from indirect lines of evidence \autocite{droop07vitamins}. For example, the amount of \bt\ required by a cell may be low and the half life of the molecules relatively long. Thus, low levels of free corrinoids do not necessarily mean that corrinoids are strongly limiting for growth. 

Direct competition experiments provide the most compelling evidence for the importance of corrinoids. I describe a competition experiment at the end of the following subsection.

\subsection{Receptor diversity and uptake arrays}

The diversity of corrinoid forms has been known for many years \autocite{roth96cobalamin}. Recent studies provide five lines of evidence about the different functional characteristics of various corrinoids \autocite{degnan14vitamin}.

First, particular bacterial species appear to require specific corrinoids for their corrinoid-dependent enzymatic reactions \autocite{yi12versatility,mok13growth,keller14exogenous}. I define a `native' corrinoid as a form required by a particular species.

Second, individual species can remodel nonnative corrinoids into their required native form \autocite{gray09the-cobinamide,yi12versatility}. \textcite{allen08identification} detected eight distinct corrinoids in human fecal samples from 20 individuals. Two individuals ingested high doses of cobalamin, the canonical \bt\ corrinoid. Those two individuals excreted broadly elevated amounts of the other detectable corrinoid forms. When those two individuals discontinued high supplementation, they reverted to a corrinoid distribution similar to the other individuals in the sample. \textcite{degnan14vitamin} suggest that the transient increase of diverse corrinoid forms in supplemented individuals implies that various bacterial gut species remodel the ingested \bt\ form into the different native corrinoids particular for each species. 

Third, many distinct corrinoid uptake receptors occur. Among 313 human gut species, \textcite{degnan14human} inferred a minimum of 27 distinct corrinoid transporter families. They used a conservative 50\% amino acid homology cutoff to define distinct transporter families. In \textit{B.\ thetaiotaomicron}, a 50\% homology cutoff did not resolve two transporter genes that had distinct \textit{in vitro} and \textit{in vivo} functional uptake properties. Using a 75\% homology cutoff, sufficient to distinguish the functionally different \textit{B.\ thetaiotaomicron} transporters, raised the estimated number of distinct corrinoid transporter families to 60. \textcite{degnan14human} further estimate that a two-fold increase in genomic sequencing would add another 43 variants, for a total of 103 among the 313 species. The specific numbers are imprecise. However, the conclusion that there are a lot of transporter variants is likely to hold.

Fourth, individual species may have multiple distinct corrinoid uptake receptors. \textcite{degnan14human} estimated an average of 1.9 distinct uptake receptor genes among 57 human gut Bacteroidetes species, with a range of 1--4 variant receptors per species. 

Fifth, competition experiments demonstrated that particular receptors provide a strong growth advantage when matched to particular corrinoids \autocite{degnan14human}. \textit{B.\ thetaiotaomicron} expresses three distinct receptors, labeled \textit{btuB1}, \textit{btuB2}, and \textit{btuB3}. An \textit{in vitro} experiment competed a mutant that expressed only \textit{btuB1} against a mutant that expressed only \textit{btuB3}. Each assay provided one of six distinct corrinoids, including cobalamin, the canonical \bt\ form. The \textit{btuB1} strain won in the presence of two corrinoid variants, whereas \textit{btuB3} won in the presence of the other four variants. 

An \textit{in vivo} experiment colonized germ-free mice with a wild-type strain that expressed all three receptors and a mutant strain with \textit{btuB2} knocked out. Mice were fed a diet that included cobalamin. The wild-type strongly outcompeted the mutant. Reintroduction of \textit{btuB2} to the mutant strain recovered nearly all of the lost competitive success. Thus, \textit{btuB2} provides a strong \textit{in vivo} competitive advantage for the uptake of cobalamin \autocite{degnan14human}. 

\subsection{Design of corrinoid receptor arrays}

Uptake of corrinoids can strongly influence competitive success and community dynamics. Three aspects seem important: corrinoids are diverse, receptors differ in their rate of uptake for different corrinoids, and corrinoids can be remodeled into native form.

Some species express more than one receptor type. How does natural selection design the array of uptake receptors expressed by such species? Here, I briefly sketch a conceptual frame for future theory and experiment.

\subsubsection{Why are there different types of corrinoid?}

\subsubsubsection{Escape from attack} Colicins and phage use corrinoid receptors as a point of attack \autocite{bradbeer76transport,di-masi73transport,kadner77relation}. Colicins are toxins secreted by bacteria that can kill other bacteria. Typically, a colicin binds to a receptor that provides an important uptake function for the target cell. The receptor function makes it difficult to hide or modify the receptor, providing a point for attack. Similarly, phage are viruses that initiate infection by binding to important cellular uptake receptors. A rare variant corrinoid and matching receptor can escape from attack by common colicins and phages. The advantage to rare variants promotes diversity \autocite{smith05evidence}.

\subsubsubsection{Prevent uptake by competitors} Production and uptake of a novel variant prevents competition for uptake by other strains \autocite{degnan14human,niehus17the-evolution}. Continual selection for novel private variants could diversify corrinoids. Initially, a strain may produce and take up its private form. When cell death releases sequestered molecules, nearby genetically related types can take up the novel form. As an initially private type gains an advantage and becomes more common, other strains may evolve uptake receptors tuned for the novel form. 

Alternatively, a private variant may arise in a cross-feeding relationship. One strain may produce a novel variant that can be taken up by another strain. If the receiving strain gains a growth benefit from uptake, and also has some positive feedback effect on the initial producer strain, then a synergistic mutualism may drive the initial spread of a private variant \autocite{frank94genetics,frank95the-origin}. Community structure may be influenced by a network of private exchange channels. Any particularly successful and abundant private line of exchange will become subject to hijacking by other species, altering the community network.

\subsubsubsection{Different biochemical properties} Particular bacteria require specific corrinoid forms \autocite{yi12versatility,mok13growth,keller14exogenous}. Thus, different corrinoids have different biochemical properties and may not be functionally interchangeable, at least in certain species. These observations lead to questions about the sequence of events in corrinoid divergence. Did the corrinoid differences arise initially to catalyze different biochemical transformations? Or did the corrinoid differences arise initially by another process, such as escape from viral attack of receptors, followed by corresponding changes in the molecules and metabolic processes that depend on corrinoids?

\subsubsubsection{Chance events and drift} Corrinoid forms may initially have changed by a sequence of minor alterations, each by itself with little effect on fitness. Over time, some changes may have led to corresponding changes in the molecules that react with the corrinoids. Evolutionary drift among lineages possibly led to diverse corrinoid forms. It seems unlikely that the total diversity of corrinoids arose by such chance events, given the strong potential fitness effects of diversity listed above. However, chance probably plays some role in aspects of diversity. 

\subsubsection{How are the properties of receptors tuned to maximize the uptake benefits of an array?}

I first summarize what is known. I then turn to various conceptual issues that frame how we may understand receptor arrays.

\textcite{degnan14human} inferred that many species of human gut Bacteroidetes express multiple corrinoid receptors. However, the binding and uptake properties of such receptor arrays for different corrinoids remain unknown, except for \textit{B.\ thetaiotaomicron}. In that species, \textcite{degnan14human} used the studies mentioned above to infer four aspects of uptake.

First, any one of the three distinct receptors takes up canonical \bt\ (cobalamin) sufficiently to confer full growth rate. In a mutant strain with all three distinct receptors knocked out, reintroduction of any one of those receptors fully restores growth \textit{in vitro} in the presence of cobalamin.

Second, although \textit{in vitro} growth by an isolated strain appeared to be invariant, direct \textit{in vitro} competition between different mutant strains revealed differences in uptake of cobalamin. Pairwise competitions were conducted between wild-type, with all three receptors, and the three mutant types that express only one of the receptors. The mutants are labeled \textit{btuB1}, \textit{btuB2}, and \textit{btuB3}, for the single receptor expressed. The growth of \textit{btuB2} was nearly identical to wild-type growth, \textit{btuB3} was mildly outcompeted by wild-type, and \textit{btuB1} was strongly outcompeted by wild-type. A further experiment showed that \textit{btuB3} outperforms \textit{btuB1} in direct competition. Thus, all three receptors can take up cobalamin, but do so with widely differing efficacies.

Third, five different corrinoids plus cobalamin were used for six direct \textit{in vitro} competitions between \textit{btuB1} and \textit{btuB3}. The receptor \textit{btuB1} won two of the six competitions, and \textit{btuB3} won the other four.

Fourth, \textit{in vivo} competition in germ-free mice confirmed that \textit{btuB2} has greatest efficacy for uptake of cobalamin. In particular, wild-type with all three receptors outcompeted a mutant with \textit{btuB2} knocked out and the other two receptors intact.

The key points are: multiple receptors exist, each receptor takes up a variety of corrinoids, and the competitive performance of each receptor varies by corrinoid type. The ability of a receptor to take up more than one corrinoid may occur because different corrinoid structures are partially constrained by their role as cofactors in particular biochemical transformations.  These points provide a basis for evaluating how different processes shape the characteristics of receptors arrays. The following paragraphs list some candidate processes. A later section discusses ways in which to approach experimental and comparative tests.

\subsubsubsection{Receptor design on a line} Begin with an overly simple case. Imagine that we can locate the potential uptake properties of each corrinoid type along a single dimension. Given the corrinoid uptake property locations along the line, how should the uptake receptor array be designed to maximize the benefit of uptake? To answer that question about receptor array design, we need to add some assumptions. 

Set constant concentrations for free corrinoids. Suppose there is a fixed cost for encoding and producing each additional receptor type. Assume there is a fixed total number of surface receptors produced, so that more of one type means less of other types. Let there be a diminishing benefit of total uptake. Locate each potential receptor along the corrinoid line. The uptake rate of a receptor for a particular corrinoid diminishes with the distance between the receptor and the corrinoid location along the line. 

With numerical values for these assumptions, we could calculate the optimal receptor array design to maximize the benefits of corrinoid uptake. An  optimal design would specify the number of different receptor types, the number of surface receptors of each type, the location of each receptor type along the line, and the best tradeoff between the maximum uptake rate and the diminishing uptake rate with distance between corrinoid and receptor. However, the purpose is not to make a specific calculation, for which we have left out many obviously crucial aspects. Instead, these assumptions begin the sketch for this complex problem. We continue to fill in the sketch and to build the conceptual frame by which we may approach this problem.

\subsubsubsection{Uptake tradeoffs} Distance along a line provides a simple way to specify tradeoffs. The closer an uptake receptor moves toward a particular corrinoid, the stronger that receptor's uptake of approaching corrinoids, and the weaker that receptor's uptake of receding corrinoids. We could increase the number of dimensions or alter the topology in which we locate receptors and corrinoids. Or we could directly specify how changes in a receptor affect the uptake tradeoffs among the set of available corrinoids. 

\subsubsubsection{Production, decay and loss} The availability of particular free corrinoids depends on the production rate by primary producers or the remodeling rate by secondary consumers. Availability also depends on the decay rate when free and when sequestered within cells, the sequestration time within cells before release by cell death, the loss to a community locale by outflow, and the gain by inflow \autocite{frank14microbial}. 

\subsubsubsection{Competition for free corrinoids} Uptake of particular corrinoids by one cellular type reduces the availability of those corrinoids for other cellular types. Thus, the rate of uptake by a cellular type and the abundance of that type may influence the growth dynamics of the other types in the community. The growth dynamics in turn influence the uptake and release of free corrinoids. 

\subsubsubsection{Fluctuation and the timescales of acquisition and use} The availability of free corrinoids likely varies over time and space. The tuning of receptor arrays will depend on those fluctuations in relation to the temporal and spatial scales of various other processes. For example, rapidly fluctuating concentrations matter little if the timescale for acquisition and internal storage within cells is relatively long. Alternatively, if corrinoids tend to be released in low frequency pulses, then rapid scavenging during those rare periods may determine success. 

Rare pulses may occur if there are causes of widespread cell death, such as viral epidemics. If rare pulses of different corrinoids happen in a relatively uncorrelated way, then cells may gain by a broad array of receptors. By contrast, highly correlated pulses may favor cellular types to concentrate on single receptors tuned to particular corrinoids. 

\subsubsubsection{Conditional versus continuous expression} Turning receptor expression on or off in response to the availability of matching corrinoids may be advantageous. However, such conditional expression may miss taking up rare pulses or otherwise fluctuating concentrations. The benefit of conditional expression depends on the availability of information about external concentrations, the cost of sensing and integrating that information, the cost of turning on expression, and the timescales of ramping up expression and turning it off relative to the frequency of corrinoid fluctuations.

\subsubsubsection{Corrinoid form in relation to uptake} Biochemical function shapes corrinoid form. The uptake properties of free corrinoids by cellular receptors may also influence their form. Consider what happens when a primary producer or remodeler of corrinoids dies and releases its molecules. If genetically related individuals of the same species take up those free corrinoids, any modification of corrinoid form that increases such uptake is favored by kin selection \autocite{hamilton72altruism,frank98foundations}. 

If other species with positive mutualistic feedbacks on the primary producer take up the corrinoids, natural selection favors increase in such uptake. If other species with negative competitive feedbacks on the primary producer take up the corrinoids, natural selection favors reduction in such uptake. Those positive and negative consequences of uptake can influence the biochemical form of corrinoids and the uptake properties of associated receptors.

\subsubsection{Theory}

Many factors influence receptor array design. Those factors and their potentially complex interactions suggest a rich subject for theoretical development. This theoretical topic generalizes the important unsolved engineering problem of multisensor array design for signal detection and estimation \autocite{bicchi94optimal,johnson15sensor}. 

Sensor design for signal detection is roughly similar to receptor design for uptake of free corrinoids. In sensor design, one assumes that uptake or measurement does not influence concentration or signal intensity. Independence with respect to uptake would approximately hold when uptake of free corrinoids is relatively small compared with other sources of outflux loss. Other sources of outflux loss may include outflow or decay or uptake by a fixed entity. Signal detection and uptake are roughly described as problems of tuning sensitivity with respect to certain costs and benefits.

The full problem of uptake receptor design for corrinoids extends the signal detection problem by allowing, in effect, competition for the signal. Uptake, or sensor measurement, by one receptor reduces the available signal for uptake or measurement by other receptors. Additionally, successful uptake leads to an increase in the abundance of the associated receptor by population dynamics and natural selection, causing a complex competitive game-like quality to the problem of uptake. 

Although this complex uptake game transcends the signal detection problem, it is useful to keep in mind the core similarities. Many aspects of the uptake problem and the sensor design problem depend on the same issues of sensitivity, tradeoffs, and fluctuations. By recognizing the abstract structure of the receptor array design problem, one can take advantage of the existing theory, develop new theory that broadens understanding for a wide range of problems, and consider uptake receptor design in biology as part of a broader subject of sensor and uptake properties of organisms. 

All of that makes for an interesting theoretical subject. But how can such theory help to understand corrinoid uptake and related problems of siderophore and glycan uptake? Before turning to that practical question, which concerns the design of experiments and comparative tests, I first summarize aspects of siderophore and glycan biology. 

Siderophore and glycan uptake share many similarities with corrinoid uptake, but also key differences. For example, glycan receptor systems have distinct molecular components for uptake and for sensory measurement of glycan availability. The greater complexity of glycan systems compared with corrinoid systems may reflect the greater diversity of glycans. In addition, the timescale for which cells must acquire glycans differs from the timescale for which cells must acquire corrinoids. The comparison of different systems provides insight into the problems of sensing, uptake, and conditional response.

\section{Siderophores}

Organisms require iron for many metabolic processes. Available iron for uptake can limit microbial growth \autocite{cornelis10ironb}. Many microbes secrete siderophores to chelate free iron. Cells take up external siderophore-iron complexes through specific surface receptors and transport mechanisms. 

Siderophores and corrinoids share certain aspects of uptake, diversity, and competitive consequences. The receptor and transport systems for siderophores and corrinoids derive from the same family of ABC transporters \autocite{cracan13metallomics}. The close homology often makes it difficult to distinguish siderophore from corrinoid transporters by amino acid sequence \autocite{yi12versatility}. 

Diverse siderophores occur. A species may take up a variety of its own secreted siderophores and various siderophores secreted by other species \autocite{cornelis02diversity,chakraborty13iron}. Individual cells may express multiple uptake receptors with different specificities. Siderophores mediate competition for uptake of free iron, which can determine the fitness of competing types \autocite{hibbing10bacterial}. For example, \textit{Streptomyces coelicolor} increases secretion of at least 12 distinct siderophores in response to competition by siderophores secreted by five related bacterial species \autocite{traxler13interspecies}.

Other studies also infer large repertoires of siderophores deployed by certain species. \textcite{baars16the-siderophore} inferred over 35 distinct metal-binding molecules secreted by the bacterium \textit{Azotobacter vinelandii}. Those molecules are primarily siderophores that bind iron, although binding of other metals occurs. \textcite{cornelis09a-survey} inferred 16 siderophore receptors in \textit{Pseudomonas syringae} and 42 in \textit{P.\ fluorescens} Pf5. Individual species can often take up siderophores produced by other species. For example, \textit{P.\ fluorescens} encodes a significant repertoire of receptors for cross-species uptake \autocite{cornelis02diversity}.

The following subsections compare four aspects of siderophore and corrinoid receptor arrays.

\subsection{Specificity of uptake receptors} 

Siderophore and corrinoid uptake may differ with regard to specificity. Siderophore receptors are often described as specific. For example, \textcite{rabsch91the-specificity} state: ``Every siderophore utilized by \textit{Escherichia coli} has its corresponding outer membrane receptor: ferric enterobactin (FepA), ferric citrate (FecA), ferrichrome (FhuA), coprogen (FhuE), aerobactin (Iut) and other catecholates (Cir and Fiu) (Hantke 1990).''\nocite{hantke90dihydroxybenzolyserine} 

Determination of specificity depends on context. For example, siderophores comprise many distinct molecular families, each family with diverse forms. Observed specificity often means that a receptor takes up one of the tested siderophore families relatively well and the other test families relatively inefficiently or not at all. Such observations do not rule out receptors that can take up closely related siderophores, perhaps with differing efficacies. I do not know of studies that have comprehensively measured the different uptake efficacies of receptors for a variety of closely related siderophores.

A corrinoid receptor can take up different forms with varying efficacy, as summarized in the prior section. However, that conclusion followed from a single detailed study of one bacterial species and a relatively small number of corrinoid forms \autocite{degnan14human}. Thus, the available studies only provide a hint that siderophore receptors may be more specific than corrinoid receptors. 

Differences in siderophore and corrinoid biology provide clues about potential differences in specificity. Siderophores are secreted to bind free iron or to take up external iron from other iron-binding molecules. Siderophore receptors may be tuned to take up native siderophores secreted by the same cell or nonnative siderophores secreted by other cells. Diversity of form may be influenced primarily by extracellular scavenging in different physical and competitive environments \autocite{kummerli14habitat}. Specificity of cellular uptake in relation to competition may favor differentiation into private types and increased diversity \autocite[see earlier section \textit{Why are there different types of corrinoid?}]{niehus17the-evolution}.

By contrast, corrinoids do not appear to be secreted. Instead, they function as cofactors for complex biochemical transformations. The special biochemical requirements of cofactor action shape corrinoid form into large, intricate molecules. Free corrinoids probably exist only through release after cell death. To the extent that corrinoid diversity may be shaped by competition for uptake, corrinoid form may be constrained by function as a cofactor and by the high expense for producing the corrinoid's intricate molecular structure. However, certain parts of a corrinoid molecule may be more easily modified than other parts, providing some opportunity for competition to shape corrinoid structure in relation to receptor binding.

Overall, the differing constraints on form and function suggest that siderophores are likely to be more diverse and have greater specificity of uptake receptors than corrinoids. Indeed, siderophores comprise a wide diversity of families with differing forms. Comparative study of the diversity and uptake specificity of siderophores and corrinoids would be valuable.

\subsection{Fluctuation and the timescales of acquisition and use}

Consider the processes that shape the spatiotemporal frequency spectrum of availability. For corrinoids, free molecules come from death of primary prokaryotic producers and release by secondary consumers. Free molecules vanish by uptake, intrinsic decay, and outflow. Fluctuations may follow rare low frequency pulses or common high frequency pulses or a mixture of different frequencies. Receptor array design will tune to the frequency pattern of availability with respect to the intrinsic cellular timescales of acquisition, use, and internal loss.

Siderophore receptor arrays face the same design issues. However, the fluctuation rhythms likely differ from corrinoids. Available iron inflows and outflows may arise from organic and inorganic sources. The diversity and abundance of various secreted siderophores mediate the competition among siderophores for chelating iron, the ways in which the physical environment shapes the flux of iron-siderophore complexes, and the competition for cellular uptake of those complexes. 

The game-like competition for iron-siderophore complexes likely shapes the frequency spectrum of fluctuations in availability in ways that differ from corrinoids. Primary production of corrinoids arises mainly for the internal use by the producers, perhaps also augmented by any benefit of released molecules that are taken up by genetic relatives or by mutualist species. By contrast, primary production of siderophores arises mainly for subsequent uptake.

If a particular siderophore is being produced only in rare pulses, then additional production may be favored because of the game-like competitive advantage for rare types associated with rare receptors \autocite{lee12an-evolutionary,inglis16presence}. Such increase in production of rare types may tend to smooth out the frequency spectrum of available siderophores. The point here is not the particular pattern that results from the game-like dynamics, which may depend on many processes, but the fact that the siderophore pattern of fluctuations likely has a stronger game-like quality than the corrinoid pattern of fluctuations. 

How should a receptor array be tuned to the fluctuations of availability across the diverse range of siderophore types? That remains an open question, one that takes us again to the broader issue of how to design arrays of sensors and uptake receptors for fluctuating inputs. There has been some discussion of this topic in relation to cellular receptors \autocite{hasty10systems}. The case of siderophores is particularly interesting because of the diversity of forms, but also is particularly complex because of the competitive, game-like nature of uptake and because production may be influenced by iron availability and siderophore concentrations.

Several experimental studies of siderophore competition have been published recently \autocite{west07the-social,hibbing10bacterial}. However, I do not know of any experimental studies that have explicitly focused on the frequency spectrum of fluctuations in iron availability and physical aspects of the environment that influence uptake. It would be interesting to consider the various costs and benefits of alternative receptor array designs under various experiment settings.

\subsection{Conditional versus continuous expression of receptors}

Cells may turn on or increase expression of receptors in response to external availability of matching types. The benefits of such conditional expression may differ between corrinoids and siderophores.

For corrinoids, the benefits of conditional expression may be limited. If cells require relatively few corrinoid molecules, then it may be more important to capture some molecules when available rather than to ramp up expression to capture a large number of molecules. If the decay rate for pulses of external availability is faster than the ramp up time for expression, then conditional expression may often miss the opportunity for capturing molecules during sporadic pulses of availability. 

By contrast, two aspects of siderophore biology may favor conditional expression of siderophore receptors. First, siderophores are actively secreted, whereas corrinoids are passively released. Thus, external availability of siderophore-iron complexes depends in part on the secretion rate. The game-like dynamics of competition shape secretion rates, which may lead to fluctuating availability. The timescale of fluctuations in secreted molecules is likely to be longer than the timescale for ramp up of conditional receptor expression, providing a potential benefit for conditionally increased expression. 

Second, the diversity of siderophore types appears to be much wider than the diversity of corrinoid types. Thus, cells may alter expression profiles of receptors in relation to the spectrum of available siderophores. Certain bacterial species appear to adjust the spectrum of their secreted siderophores in relation to specific competitors \autocite{traxler13interspecies}. When such conditional adjustments in secretions occur, associated adjustments in receptor expression seem likely.

Various physical and external aspects also influence the relevant timescales for available iron and siderophores. Diffusion rates and extrinsic causes of inflow and outflow affect fluctuations and availability. The rates at which cells can sense and alter receptor expression profiles must be considered in relation to the rates of change in availability.

There are few observations about conditional expression of siderophore receptors \autocite{sexton17pseudomonas}. I presented these speculative comments to stimulate the collection of basic data and the development of new theory.

\textcite{dumas13switching} presented an interesting study in which \textit{Pseudomonas aeruginosa} switches its siderophore secretion between alternative types. Severe iron limitation triggers expression of the highly efficient iron scavenging pyoverdine siderophore. As iron availability increases, expression switches to the less efficient pyochelin siderophore. Pyoverdine is metabolically more costly to produce than pyochelin. Apparently, increasing iron availability causes \textit{P.\ aeruginosa} to switch from the relatively efficient and costly pyoverdine to the relatively inefficient and cheap pyochelin. This study did not address conditional expression of receptors. However, it did show how a species may alter its siderophore expression in response to changing external conditions. 

\subsection{Competition and inverse public goods}

Are cells simply foraging for their growth requirements? Or do cells increase receptor expression to take up excess molecules and competitively reduce the growth of their neighbors?

Reducing availability for competitors is an inverse public goods problem. Siderophores are a public good \autocite{west07the-social}. Once secreted, they become publicly available for uptake by any cell with an appropriate receptor. Excess uptake of siderophores reduces availability of a public good. The shaping of inverse public goods uptake by natural selection has several aspects. 

First, a cell that takes up excess molecules reduces availability for genetic relatives and for competitors. If the harm to genetic relatives is less than the harm to competitors, then the trait may be favored.

Second, excess uptake provides benefits to a genotype in relation to the abundance of that genotype. When rare, excess uptake by a clone has little effect on availability for competitors. The costs of excess uptake may be relatively insensitive to abundance, including the costs of maintaining internal iron homeostasis \autocite{andrews03bacterial}. Thus, rare types may lose more than they gain by excess uptake. When abundant, excess uptake by a clone may significantly reduce availability for competitors. If cells in a clone can take up excess molecules in a way that reduces competitor growth more than it reduces clone-mate growth, then inverse public goods traits can be favored.

Overall, competition will depend on interactions between abundance of cellular types, genetic relatedness between cellular types, availability of free molecules for uptake, rates of uptake by different genotypes, costs of excess uptake, and the recycle rate of molecules taken up in excess. 

Interference of iron availability occurs as a host strategy to control pathogens \autocite{cherayil11the-role}. Mammals sometimes increase their sequestration of iron in response to infection, reducing the available iron for invading pathogens. Thus, interference competition over iron is plausible. However, the host-pathogen situation differs from interference competition between different microbes.

Competitive excess uptake may occur differently for siderophores and corrinoids. In siderophores, the diversity and specificity of types may be greater than for corrinoids. If so, excess competitive uptake of siderophores may benefit more strongly from a greater range of expressed receptor types rather than a greater level of expression for particular receptors. By contrast, excess competitive uptake of corrinoids may be more strongly influenced by upregulating the level of expression for a few receptor types.

\section{Challenges for uptake receptor array design}

This section briefly summarizes the key challenges of uptake receptor array design for corrinoids and siderophores. Those challenges provide focus for the next example of glycans.

Corrinoid uptake presents a relatively simple challenge. A cell needs a small number of corrinoid molecules. A cell can take up its preferred type of corrinoid. Or it can take up a variant corrinoid, which can either be used directly or remodeled into usable form. 

I discussed a variety of design issues for corrinoid uptake, including the number, specificity, and conditional expression of receptors. I emphasized that array design likely depends on the frequency spectrum for fluctuations of external availability. The availability spectrum must be considered relative to various cellular timescales. Potentially important timescales include decay of internal molecules and ramp up time for conditional increase of receptor expression.

Siderophore uptake presents a challenge of intermediate complexity. A cell may obtain necessary iron by uptake of iron-siderophore complexes. Siderophores primarily function to chelate iron and bind to cellular receptors for uptake. Siderophores are more diverse and perhaps more specific with respect to uptake receptor binding than corrinoids. That greater diversity and specificity likely arises because siderophores are primarily designed for uptake, whereas corrinoids are primarily designed for their biochemical function as cofactors for complex biochemical transformations. 

Siderophore's game-like dynamics for external iron scavenging, receptor binding, secretion rates, and fluctuating abundances complicate the challenges of uptake receptor array design. Cells often take up a variety of siderophores produced by other species in addition to their own native forms. The frequency spectrum for fluctuations of external availability sets the key challenge, as for corrinoids. The timescales of fluctuating availability must be measured against the intrinsic cellular timescales for conditionally altering the deployment of various receptors and for varying production of native siderophores. 

The following section discusses receptor array design for glycans. Those complex carbohydrates form diverse biochemical structures. The challenges for glycan uptake differ from the previous examples. For corrinoids, cells need relatively small amounts, which are not consumed internally. Different corrinoids share basic structure, with relatively modest variety around the basic form. For siderophores, cells need a modest amount of iron. The diversity of siderophore types arises primarily through game-like competitive dynamics. 

By contrast with corrinoids or iron, the glycans themselves are structurally very diverse. Cells need a continuous supply of external energy which, for some species, comes primarily in the form of diverse glycans. The binding and initial breakdown of glycans are relatively complex challenges. The next section considers those challenges of glycan uptake in relation to the previous examples.

\section{Glycans}

Some bacteria feed only on simple sugars. Others can break down the most recalcitrant glycan fibers, such as rhamnogalacturonan II, a significant cell wall component of some fruits \autocite{ndeh17complex}. I focus on the Bacteroidetes group, which includes well studied systems of glycan uptake and digestion. I limit my discussion to a few specific challenges of receptor array design, broadening my prior discussions of corrinoids and siderophores. Many excellent reviews of glycan catabolism have been published recently \autocite[e.g.,][]{hamaker14a-perspective,martens14the-devil,grondin17polysaccharide}. 

\subsection{Receptor diversity and specificity}

Each Bacteroidetes Polysaccharide Utilization Locus (PUL) comprises multiple co-regulated genes. A PUL provides functions to acquire and digest complex carbohydrates. Among \textit{Bacteroides thetaiotaomicron}, \textit{B.~ovatus} and \textit{B.~cellulosilyticus} WH2, each has approximately 100 PULs. The set of PULs differs significantly between species pairs \autocite{grondin17polysaccharide}. \textit{B.~thetaiotaomicron} and \textit{B.~ovatus} devote approximately $18\%$ of their genomes to PULs \autocite{martens11recognition}.

Specificity may be inferred in various ways. Upregulation of particular PULs in response to certain glycans indicates specificity \autocite{martens11recognition,kabisch14functional}. Binding affinity of surface receptors affects specificity. A PUL also encodes several components that capture and digest glycans and that alter the PUL expression level. Functional specificity of digestion and response may depend on the binding specificities of the different components. 

One or more surface molecules initially capture external glycans. Cell surface glycoside hydrolases often bind and partially digest the glycan before transport across the outer membrane. Transporter molecules may influence rate of uptake and functional specificity of the PUL. Once across the outer membrane, further enzymes bind and digest the glycan components in the periplasmic space between the outer and inner membranes. PULs typically encode carbohydrate sensors and transcription factor regulators that control expression level. Individual PUL components may bind a variety of glycans or be highly specific. Overall, the different steps of binding, digestion, transport, and sensing combine into a series of filters that enhances specificity \autocite{cuskin12how-nature,gilbert13advances}. Binding specificity of individual components is an active area of research \autocite{grant14recent,tauzin16molecular,glenwright17structural,ndeh17complex}.

Surface glycan binding proteins from different PULs may interact. A binding protein may initially capture a glycan for which its PUL lacks matching digestive enzymes. By holding the glycan near the cell surface, the binding protein of another PUL may increase its capture rate. If that sort of interaction does occur, then the binding properties of surface molecules and the expression of receptor arrays may be affected.

\subsection{Receptors for partially digested components}

Large carbohydrates initially bind to cell surface receptors. Cell surface enzymes partially digest those large molecules before transport across the outer cell membrane. Some of the partially digested components may release before transport. 

\textit{Bacteroides ovatus} releases partially digested components of various xylans \autocite{rogowski15glycan}. In one experiment, wild-type cells grew on the complexly structured corn glucuronoarabinoxylan. Knockout of a single extracellular glycoside hydrolase prevented growth. That particular enzyme digests the full corn carbohydrate into relatively complex components. Culture with wild-type cells allowed the mutant to grow. Apparently, the wild-type released partially digested components that could be processed by the mutant. This study suggests that partially digested components, when released, may be taken up by the same cell or by other cells of the same species. 

\textcite{rogowski15glycan} also studied uptake of partially digested components by another species. The relatively simple carbohydrates wheat arabinoxylan and birch glucuronoxylan supported growth of \textit{B. ovatus} but not \textit{Bifidobacterium adolescentis}. Co-culture of these species allowed growth of \textit{B. adolescentis}, suggesting release of partially digested components by \textit{B. ovatus}. 

Receptor arrays may be partly designed to take up components released by the same cell, by other cells of the same species, or by other species. Release of partially digested components provides benefits to other cells, possibly of a different species \autocite{pande17bacterial}. Is extracellular binding and digestion partly designed to provide cross-feeding benefits to other species? Or is release of partially digested components simply a neutral consequence of the selfish processing of complex food sources?

\textcite{rakoff-nahoum16the-evolution} presented an example of mutually beneficial cross-feeding. \textit{Bacteroides ovatus} (\Bo) releases partially digested components that enhance the growth of a partner species, \textit{B.\ vulgates} (\Bv). In turn, an increase in \Bv\ enhances the growth of \Bo. 

The details provide insight into design aspects of binding, uptake, and processing of food sources. \Bo\ produces two cell surface glycoside hydrolases that partially digest inulin, a complex structure of fructose polymers. Knockouts of those two surface enzymes had either no detectable effect on growth of \Bo\ or, under certain conditions, improved growth. When cultured by itself, \Bo\ appears to grow best by taking up the full inulin molecule rather than the partially digested components produced by these two surface enzymes.  

The partially digested inulin components released by \Bo\ enhance the growth of a partner species, \Bv. In co-culture, the partner \Bv\ grew significantly better with wild-type \Bo\ than with a knockout that expresses only one of the surface enzymes. Presumably, the \Bo\ knockout released fewer partially digested inulin components, decreasing the growth of the \Bv\ partner. 

How does \Bo\ itself gain a benefit by releasing partially digested components that aid the partner \Bv? In a co-culture of \Bo\ wild-type and a mutant \Bo\ with both surface enzymes knocked out, addition of the partner \Bv\ enhanced growth of wild-type \Bo\ relative to the knockout. Release of partially digested components by wild-type \Bo\ cells enhanced growth of particular \Bv\ cells that, in turn, benefitted preferentially the wild-type rather than the mutant \Bo\ cells. 

A return evolutionary benefit to \Bo\ occurs only if the partner enhances growth of the \Bo\ genotype that released the components \autocite{frank94genetics}. How might such preferential return benefit occur? Spatial association is possible. The partially digested components released by wild-type \Bo\ may enhance growth of neighboring \Bv\ cells. Those more vigorous \Bv\ cells may return a benefit to their neighboring \Bo\ cells, which include a disproportionate share of the cells that initially released the partially digested components. In general, mutually beneficial cross feeding relationships provide a fascinating aspect of uptake receptor design. However, it is challenging to obtain evidence of return benefits directly to the genotype that initially releases partially digested components. 

Release of partially digested components often arises as an inevitable consequence of leakage at the cell surface by cells tuned to maximize their own selfish growth. \textcite{rakoff-nahoum16the-evolution} showed that \textit{B.\ thetaiotaomicron} cells release partially digested components of the glycans amylopectin and levan. Knockouts of the surface glycoside hydrolases for each glycan gained when grown with wild-type, indicating mutant uptake of partially digested components released by wild-type. In direct competition, wild-type outcompeted the knockouts, suggesting that extracellular digestion provides a direct benefit to the cell. Any indirect benefit to neighboring mutants is simply a consequence of leaking partially digested components that the cell fails to take up.  

Another study showed that \textit{B.\ thetaiotaomicron} limits its release of partially digested components of the yeast cell wall glycan $\Ga$-mannan \autocite{cuskin15human}. Extracellular digestion makes large oligosaccharides, almost all of which are taken up by the cell. 

The observed variation in the leakage of partially digested components raises the problem of how cells tune their processes of digestion and uptake. In some cases, tuning emphasizes direct uptake and growth benefits to the cell. In other cases, tuning may include indirect benefits gained by releasing components to aid the growth of other cells. 

\subsection{Sensors and conditional response}

A Bacteroidetes PUL typically expresses its components at a low level. A PUL-specific sensor detects particular carbohydrates and upregulates PUL expression. Cells exposed to mixtures of carbohydrates prioritize upregulated expression of some PULs over others. Hierarchical regulation differs between species, causing distinct carbohydrate utilization patterns \autocite{grondin17polysaccharide}. I briefly describe a few examples. I then discuss the general problem of how cells perceive and respond to various environments.

\textit{B.\ thetaiotaomicron} detects different carbohydrates in various ways. For levan, a complex fructose polymer, cells upregulate expression of the levan-specific PUL in response to the simple fructose components \autocite{sonnenburg10specificity}. The levan-specific PUL includes the various functions needed to transform external levan polymers to simple intracellular fructose. Initially, an outer membrane surface enzyme breaks levan into large pieces. Outer membrane receptor and transport components move the polysaccharide pieces into the periplasmic space between the outer and inner membranes. Additional enzymes in the periplasmic space digest the polysaccharides into simple fructose sugars. A surface molecule on the inner membrane senses free fructose in the periplasmic space and triggers upregulation of the PUL. Another surface molecule on the inner membrane transports fructose into the cell.

For arabinan, a complex arabinose polymer, cells detect and integrate three distinct signals that regulate expression \autocite{schwalm16multiple}. First, outer membrane surface molecules break arabinan into oligosaccharide fragments and transport those fragments into the periplasmic space. A surface molecule on the inner membrane senses those oligosaccharides and stimulates expression of the arabinan utilization locus. Separately, those oligosaccharides are broken down and transported into the cell, where they become free arabinose sugars. 

Second, an intracellular sensor molecule detects arabinose. That internal arabinose sensor is part of a distinct arabinose utilization locus.  Stimulation of the internal sensor by arabinose represses expression of both the arabinan and the arabinose utilization loci. 

Third, a distinct internal molecule upregulates expression of both the arabinan and the arabinose utilization loci. The signal that stimulates this third regulator is not known. It could be part of the system that regulates the preference for different food sources when multiple carbohydrates are present.

In summary, free fructose sugars stimulate expression of the levan utilization locus, whereas free arabinose represses expression of the arabinan utilization locus. Cells apparently perceive and respond to distinct carbohydrates in different ways.

\subsection{Hierarchy of preferred types}

Bacteroidetes species upregulate specific PULs in response to particular carbohydrates. Mixtures of carbohydrates do not stimulate all matching PULs. Instead, cells prioritize the usage of some carbohydrates over others \autocite{martens11recognition,rogers13dynamic,ravcheev13polysaccharides,pudlo15symbiotic,cao16cis-encoded,comstock16small}. 

\textcite{schwalm17prioritization} grew \textit{B.\ thetaiotaomicron} in a mixture of the polysaccharides chondroitin sulfate and arabinan. The presence of chondroitin sulfate repressed expression of the arabinan PUL. Cells first consumed chondroitin sulfate, then upregulated expression of the arabinan PUL and switched to using that food source. 

Increased expression of the arabinan PUL requires stimulation of a periplasmic sensor by arabinose oligosaccharides, as described in the prior section. The presence of chondroitin sulfate repressed phosphorylation of the periplasmic sensor by arabinose oligosaccharides. Repressed phosphorylation prevents signal transduction and increased expression of the arabinan PUL.

\textcite{schwalm17prioritization} analyzed \textit{B.\ thetaiotaomicron} growth on 55 pairwise combinations of 11 polysaccharides. They observed an overall tendency first to consume polymers not composed of arabinose or fructose sugars, followed by polymers of those sugars. In some pairs, \textit{B.\ thetaiotaomicron} preferred neither polysaccharide over the other, consuming both polysaccharides during growth. 

When paired with arabinan, the preferred polysaccharides chondroitin sulfate, pectic galactan, polygalacturonic acid and rhamnogalacturonan I repressed expression of the arabinan PUL. Presumably, depletion of the preferred polysaccharide repressed the preferred PUL and upregulated the PUL for arabinan. 

\textcite{lynch12prioritization} demonstrated the preference of \textit{B.\ thetaiotaomicron} for arabinan relative to a host mucin, a glycoprotein component of intestinal mucus. Stimulation of the arabinose oligosaccharide sensor of the arabinan PUL repressed expression of the mucin-related PUL. Thus, \textit{B.\ thetaiotaomicron} prefers various polysaccharides over arabinan and prefers arabinan over certain host mucins. Each preference associates with transcriptional repression of the nonpreferred source, mediated in this case by the same arabinose oligosaccharide sensor of the arabinan PUL.

\textit{B.\ thetaiotaomicron} prefers a variety of dietary carbohydrates over host mucins. The dietary carbohydrates repress expression of the mucin PULs \autocite{rogers13dynamic,pudlo15symbiotic}. By contrast, \textit{B.\ fragilis} and \textit{B.\ massiliensis} often prefer mucins over the limited range of dietary carbohydrates that they can digest. \textit{B.\ massiliensis} mostly expresses the opposite preferences from \textit{B.\ thetaiotaomicron} when presented with starch paired with various mucins. These related species share a common starch PUL and perhaps also components of the regulatory system that controls hierarchical expression. Nevertheless, these species have evolved different preferences for various carbohydrates.

\subsection{Perception and response}

Some bacteria can digest diverse carbohydrates. To perceive the availability of various carbohydrates, many separate subsystems must be expressed at a low level. For example, \textit{B.\ thetaiotaomicron} may not be able to detect arabinan without expressing a low level of the arabinan-specific PUL. That PUL includes the extracellular binding, digestion, transport, and sensor that may be needed to detect arabinan availability.

Cells must integrate the various signals of changing availability for glycans, partially digested components, or simple sugars. \textit{B.\ thetaiotaomicron} has a large sensor array. Its various PULs encode 32 distinct hybrid two-component sensors \autocite{sonnenburg06a-hybrid}. Those inner membrane sensors detect specific carbohydrate availability in the periplasmic space and transduce a signal into the cell. \textit{B.\ thetaiotaomicron}'s PULs also encode numerous carbohydrate sensors of the $\Gs/$anti-$\Gs$ factor family and some additional susR family sensors. How do cells integrate all of those inputs to classify the state of the environment?

For cells, classification effectively means the way in which perceived information transforms regulatory control pathways to alter the expression of various subsystems. For glycans, how do particular inputs regulate hierarchical expression of the PUL subsystems? The inputs include the various carbohydrate sensors. Other inputs likely play an important role. For example, cells may have sensors for cellular growth rate and for temporally fluctuating signals correlated with food availability. Those additional sensors may link into the regulatory control of carbohydrate usage.

Bacteria vary greatly in the diversity of carbohydrates that each species can take up and digest. That variability provides great opportunity for comparison. How does the sensory array scale with the breadth of carbohydrate food sources? What sorts of different environmental signals and sensors flow into the regulatory control of carbohydrate usage? How do the timescales of signal fluctuations shape the design of sensors and regulatory control? Many reviews summarize current knowledge of preferential resource use in bacteria \autocite{deutscher08the-mechanisms,gorke08carbon,chubukov14coordination,beisel16rethinking}.

We may ask similar comparative questions about the sensing and regulation of corrinoid and siderophore uptake. How are receptor uptake arrays designed in relation to the diversity of inputs? How do fluctuations in inputs influence design? How do timescales of uptake, usage, and need influence design? What sorts of correlated signals provide information about external availability and internal state? When does it pay to express receptors at a constant level rather than adjust in response to perceived changes in availability? Previous studies of uptake provide a basis for these more advanced questions \autocite{degnan14human,schalk16an-overview}.

\section{Deep learning and control theory}

When the availability of a particular nutrient rises, cells may increase the receptors for taking up that nutrient. However, that simple stimulus-response notion ignores many aspects of cellular biology and the temporal structure of fluctuating availability. I discussed those additional aspects in previous sections. But I did not offer a comprehensive conceptual framework in which to analyze the problem. How can we frame the problem to highlight the essential features and provide structure for future analysis?

Theory from several different fields relates to these issues. In evolutionary biology, phenotypic plasticity considers how natural selection shapes an organism's response to changes in its environment. In computer science, deep learning studies how to classify inputs and how to respond in order to achieve a specified goal. In engineering, control theory analyzes how various input signals should be transformed into actions that minimize some measure of distance between system output and an optimal goal. The Appendix briefly summarizes the relevant literatures.

In all cases, the essential problem concerns two transformations. First, sensory input leads to a classification of the environmental state. Second, the inferred environmental state leads to an appropriate response. 

\subsection{Deep learning}

Classification and response define the two primary challenges of deep learning research. Computer vision and voice recognition transform inputs into a best guess for a matching object among a set of possibilities---a classification that maps various inputs to particular outputs. Classification may be the final goal. Or, classification may provide the basis for responding to the environment. For example, when the system classifies the visual input as an increasing deviation from a target, then the system may adjust its future trajectory to be closer to the target.

Recent breakthroughs in computation and machine learning pervade modern life. New computational classification and response systems often outperform humans. The new concepts and methods comprise \textit{deep learning}. The \textit{learning} simply means using data, or past experience, to improve stimulus-response performance. The \textit{deep} qualifier refers to the computational method that triggered the revolutionary advances in performance \autocite{nielsen15neural,goodfellow16deep}. 

A deep learning system is a computational network loosely modeled after a biological neural network. A set of nodes takes inputs from the environment. Each input node connects to another set of nodes. Those intermediate nodes combine their inputs to produce an output that connects to yet another set of nodes, and so on. The final nodes produce an action. That action classifies the environmental state of the initial inputs or takes an action based on that classification. A deep neural network has many layers of nodes between initial inputs and final outputs. 

Does deep learning research provide insight into the design of bacterial receptor arrays for the uptake of nutrients? That remains an open question. Several articles have linked problems of cellular perception and response to deep learning (see Appendix). So far, few specific insights have followed. However, analogies between engineering and biological design often provide mutually beneficial insight. Such insight clarifies both the similarities and the differences between human-designed and naturally designed systems.  

I illustrate the problem with the following example. Suppose gut bacteria receive inputs associated with wine consumption. Is that wine a final evening glass, to be followed by many hours of fasting? Or is that the first glass before a multicourse dinner that will continue over several hours? 

A bacterium's capacity to distinguish and respond to those alternatives depends on the design of its regulatory control network. I discuss four aspects of network design. Those aspects include the key design features of modern deep learning networks.

\subsubsubsection{Architecture} Network structure describes the pattern of connections between inputs and outputs. A simple feedforward architecture flows unidirectionally from inputs to internal nodes to outputs. In that case, inputs arise from sensors that detect carbohydrates and outputs regulate expression of the matching receptors. Feedback loops reverse directionality. For example, greater receptor expression may repress the associated sensors, reducing sensitivity to the signal and limiting further increase in receptor expression. 

Feedback loops and other simple network motifs have been studied extensively in biology \autocite{alon07network}. Such motifs can be analyzed when they occur as simple connections between a few nodes. However, some bacterial regulatory networks may have a hundred or more input sensors. For example, \textit{B.\ thetaiotaomicron} has dozens and perhaps a hundred carbohydrate sensors. In addition, many other external sensors of the environment and internal sensors of cellular state may produce inputs into the regulatory control of the glycan uptake receptor array. This large regulatory network likely has many connections. 

Deep learning analyzes large, highly connected networks. In such networks, it is often difficult to interpret or assign significant causality to particular connections. Properties such as feedback may be present in a network. But such properties may be diffuse aspects of network architecture rather than simple aspects that can be traced between a few nodes. 

In deep learning networks, recurrence generalizes the notion of feedbacks. Roughly speaking, recurrence means that the state of a node depends on both current inputs and previous inputs. Feedback is one way to flow an older input to a node. Recurrence broadly includes the variety of diffuse architectural ways in which a network can remember and combine past inputs to drive the state of various internal nodes and outputs. 

Although recurrence is a simple principle, designing large recurrent network architectures that perform well is an ongoing challenge in the computational study of deep learning. Some work on biological neural networks also addresses this issue. 

The design of bacterial networks to regulate receptor arrays faces similar issues of recurrence. Surely, past inputs can be important in distinguishing the final night's glass of wine from the first glass that starts a large meal. A bacterial network's degree and form of recurrence likely depends on the breadth of different carbohydrates that it can take up. To distinguish between a few alternative carbohydrate food sources, simple designs, such as direct feedback loops, often work well. But for broad receptor arrays and fluctuating environments, more diffuse networks of the types in deep learning may perform better. 

The goal in biology is to understand how the particular structure of the external challenge corresponds to the variety of network architectures. By considering the various challenges of glycan, siderophore, and corrinoid uptake, we gain a broader perspective on the role of network architecture.

\subsubsubsection{Representation and classification} A well designed network must represent the environmental inputs and system outputs in an effective way. \textit{Effective} often means discriminating between likely alternatives that matter for performance. In our wine example, the alternatives concern whether the wine will be followed by a large meal or a long nighttime fast. Sensors only for simple sugars may not allow an accurate representation of attributes such as wine consumption, prior fasting period, and time of day. Instead, the sensors must be tuned to build a representation of the problem from meaningful parts. 

A representational part might be a component of wine. Another part might correlate with time of day, responding to molecules that follow a circadian rhythm. Yet another part might pick up cues about time since the last meal. A cue about a time interval may combine a sequence of previous inputs in a recurrent manner. 

Those high-level concepts may correspond to an internal network node that takes inputs from many individual sensors or from other internal nodes. For example, the individual sensors may detect various simple sugars and complex polysaccharides. The profile of sugars and polysaccharides may be combined together to represent particular types of consumption, such as wine versus nonalcoholic fruits. 

As information flows from low-level sensors to deeper internal nodes, the representations may take on higher-level aspects of the alternative environment states, such as wine, time  of day, and prior fasting period. At that high level point, the network has effectively classified broad aspects of environmental state.

For glycans, the actual sensors, representations, and classifications will of course depend on the particular organism, its environment, and the costs and benefits of building a network to link inputs with outputs. The point here does not concern wine and meals, but the fact that we can think about regulatory control with respect to sensors, representation, and classification. These principles define many of the current challenges and fascinating progress in the study of deep learning.

Comparatively, it is useful to think about the different kinds of challenges that arise for glycans, siderophores, and corrinoids. How do those challenges lead to particular architectures, representations, and classifications? 

\subsubsubsection{Response} Sometimes, simple classification corresponds closely to the ultimate goal. For example, an environmental state that strongly correlates with an upcoming nighttime fast may favor reduced expression of uptake receptors and digestive enzymes. The cell only needs to classify the environment in terms of likelihood of an upcoming fast. 

Often, classification and response cannot be separated. For example, an upcoming nighttime fast may describe one attribute of the environment. But responses to that environmental classification may vary depending on other environmental attributes, such as availability of host mucins for digestion. We can think of mucins as another attribute of classification. However, every environmental aspect is potentially another attribute of classification. A meaningful classification gains its meaning with respect to potential alternative responses. 

We may combine classification and response in the following way. Consider the space of inputs over which the best response remains invariant. Then we can partition the environment into subsets, each with invariant responses. In practice, we want the best cost-benefit tradeoff for sensors that pick up the environmental correlates and that allow estimation of which response-invariant partition includes the current environment. 

That notion of partitioning the sample space matches the essence of finding sufficient statistics in statistical theory. The concept is relatively simple. Yet making a network that performs with such sufficiency and invariance is far from trivial in practice. The difficulty of applying the concepts in practice brings us back to architecture, representation, and classification. Deep learning develops the methods that make the simple overarching principles work in application. That practical success suggests that there are simple underlying concepts that remain to be fully understood.

\subsection{Control theory and the frequency domain}

Return to the basic problem. A nutrient type comes in different forms. Availability fluctuates. How should cells design receptor arrays to capture the various nutrient forms? When does it pay to gather information with sensors and adjust receptor deployment? When does it pay to use a small static set of receptors, without adjustment to changing availability?

Deep learning focuses attention on network architecture, representation, classification, and response. That perspective separates the design problem into key components. Separation provides a clear way to parse observable patterns and to develop theory and analysis.

Control theory provides a complementary perspective. A control system transforms environmental signals into action signals. For example, signals of environmental temperature may be transformed into signals that raise or lower heat output \autocite{suykens96artificial,ogata09modern}.

Control theory emphasizes signal transformation. We can think of a receptor array as transforming signals of external nutrient fluctuations into signals of internal nutrient flow into the cell. Similarly, we can think of nutrient sensors as transforming signals of nutrient fluctuations into internal signals that control receptor deployment. 

Consider a daily fluctuation of an available nutrient. One can think of the rise and fall of the nutrient with time, repeating the cycle in each daily period. How does a sensor transform the external signal into an internal signal? If the sensor integrates information over a daily period, then the internal signal will be constant. If the sensor integrates information over a few hours, then the internal signal rises and falls each day, but with a shifted peak and lower amplitude. 

Those examples describe the external and internal signal fluctuations over time. For very simple fluctuation patterns, it is relatively easy to track the changes in the time domain. In reality, actual fluctuations usually combine many different processes acting over different periods. Rare low frequency bursts arise by occasional epidemics, changes in weather, and so on. Common high frequency fluctuations arise by stochastic processes of death, local flow, and so on. Fluctuation patterns combine those various frequencies.

One can think of a sensor or an uptake receptor as a filter. The filter passes certain types of external fluctuations through as internal signals and blocks other types of fluctuations. For example, rapid high frequency fluctuations may happen too quickly to change the rate of uptake by a receptor or the internal signal passed through by a sensor. By contrast, low frequency changes may pass through such filters. 

These examples show that, instead of thinking about fluctuations as changes through time, we can think about them as happening with certain frequencies. Each frequency of fluctuation has a corresponding intensity or power. A receptor array, as a control system, transforms the frequency power spectrum of external signals into a spectrum of internal signals. Those internal signals are, in turn, transformed into a spectrum of action signals. 

Fluctuating signals measured over time contain the same information as fluctuating signals measured by the frequency power spectrum. But, for complexly fluctuating signals, it is much easier to think about how systems transform signals in terms of the changes in the frequency power spectrum.

These descriptions of control theory and frequency domain analysis are widely used in both engineering and systems biology. However, once the equations and technical aspects begin, it is easy to lose sight of the simple underlying qualitative principles. Those simple principles can help greatly in the formation of testable predictions and in the design of experiments to test those predictions.

\section{Discussion}

I highlight key processes, grouped into five topics. Ideally, each process associates with specific predictions and empirical tests. However, predictions and tests require a clear conceptual structure and some basic facts. The concepts and facts are not yet sufficient for many aspects. I develop a first draft, which leaves gaps. Filling those gaps will advance the subject.

\subsection{Diversity of available nutrients and matching receptors} 

Diversity arises by divergence of an ancestral type into two. What are the key process of divergence? 

\process{Function} Different corrinoid forms may catalyze different biochemical reactions. Different siderophore forms may bind iron differently. Do the molecular changes that alter function also influence receptor uptake? If so, then functional divergence may lead to receptor array diversity, whereas functional constraints may limit receptor diversity.

Most glycans function independently of the bacteria that take them up. Mammalian mucins that line the gut may be an exception. If gut mucins were simple and lacked diversity, then bacteria could easily digest through the mucosal protective layer by using just a few matching mucin utilization gene clusters. However, gut mucins are often highly complex and diverse. That complexity and diversity prevent bacteria from attacking by use of a few simple gene clusters.

\process{Neutrality} Corrinoid and siderophore forms may initially change by minor alterations with little effect on function. Over time, corresponding changes may arise in matching receptors and in other molecules that interact functionally with the altered forms. Glycans usually function independently of bacterial consumers.  Thus, glycan diversity can be regarded as fixed by external processes. 

\process{Escape} Toxins and viruses target common receptors for entry into cells. Rare receptor variants escape attack. Rare-type advantage favors diversification of receptors. Diversification of receptors may drive diversification of corrinoids and siderophores. For glycans, receptor escape may favor uptake of similar glycans by variant receptors or diversification of glycan usage between species. 

\process{Privacy} Genotypes that produce and take up novel variants can escape competition. The potential benefits of private variants differ by nutrient type \autocite{degnan14human,niehus17the-evolution}. 

Corrinoid private forms, released at cell death, benefit nearby genetic relatives. A novel variant may also become a private channel to send benefits to mutualistic partners. If the private benefit enhances the growth of a partner genotype or species, then the enhanced partner growth may return a benefit to the original producing genotype.

Siderophores are actively secreted. A private form could be taken up by the secreting cell or its genetic relatives, providing a direct return benefit. Or a private variant could enhance the growth of a partner in a mutualistic cross-feeding relationship. 

Glycan forms arise extrinsically. Cells may partially digest glycans into novel or rare components. Such private components may be taken up by genetic relatives or by cross-feeding mutualistic partners. 

\process{Predictions} A few examples illustrate the potential. For function, a weaker correlation between changes in function and changes in receptor binding may lead to greater diversity. Controlled mutational changes to corrinoids and siderophores could allow measurement of the function-binding correlation. Experimental evolution studies could select for novel function and examine the correlated response in receptor binding.

For escape, highly abundant species may be under more intense selection to vary their receptors than rare species. If so, one expects greater receptor diversity in abundant species. Experimental evolution could vary attack rates for various receptors and study the patterns of receptor diversification and associated functional changes. 

Privacy in common species may drive sequential evolution of new variants. As each new private form rises in abundance, competitors may evolve to take up that form. By contrast, rare species may be more likely to maintain private variants without generating as much sequential novelty. Siderophores may more easily develop private variants than corrinoids, because of the relatively large, costly and complex aspects of corrinoid function. 

A greater correlation between function and receptor binding should limit relative opportunities for generating private variants. Mutualistic cross feeding by private variants may arise more easily in spatially structured environments that allow partners to return benefits to the producing genotype \autocite{pande16privatization}. 

Environmental aspects of resource abundance and nutrient flux rates may influence diversity. For example, resource rich and iron poor environments may favor greater siderophore diversity than resource poor and iron rich environments. 

\subsection{Number and specificity of receptor variants per cell} 

This section discusses uptake processes that influence receptor diversity.

\process{Separation} A single receptor may be able to take up variant forms of a nutrient. The potential to take up variants depends on how different the variants are. That separation between nutrient variants arises by the processes that cause divergence. Processes such as altered function or neutrality drive divergence for reasons that have nothing to do with receptor binding. Those processes of divergence may not cause a large separation with regard to uptake, allowing a single receptor to take up more than one variant. By contrast, escape and privacy drive divergence in order to separate variants with regard to receptor binding. In that case, each receptor likely takes up only one variant.  

\process{Tradeoffs} For a given receptor, increased uptake of a particular nutrient may reduce uptake of variant forms. Such uptake versus specificity tradeoffs influence the design of individual receptors and the design of receptor arrays. Tradeoffs between binding affinity and specificity are well known in other systems, such as in antibodies and other aspects of vertebrate immunity \autocite{frank02immunology}.

Individual receptors may be tuned to low affinity and broad specificity. That broad tuning allows a few receptors to take up a relatively wide set of variants. Alternatively, each receptor in an array may be maximally tuned for high uptake rate of a particular variant. That narrow tuning requires more receptors to take up a wide set of variants. The tuning of each receptor depends on the separation between the nutrient forms and on the tradeoff between uptake rate and specificity. If the expression of an additional receptor is costly, then it may pay to locate a single receptor between two separated nutrient variants. That midpoint tuning provides a low rate of uptake for both variants.

\process{Filtering} I have used the word `receptor' for a variety of distinct uptake processes. Uptake processes can often be thought of as sequential filters. For glycans, outer membrane molecules initially bind various external polysaccharides. Surface enzymes catalyze partial digestion of specific polysaccharides. Of those polysaccharides that are partially digested, the resulting components are transported into the periplasmic space. Further digestion occurs, followed by transport of sugars and simple polysaccharides through the inner membrane. Each step filters the initial cell-surface binding process, altering the rate and specificity of uptake. In some cases, the initial surface digestion process may release partially digested components that are taken up by other receptors.

Binding and uptake may be simpler for corrinoids and siderophores than for glycans. However, some multistep filtering does occur in the simpler cases. For example, certain bacteria take up corrinoids that differ from the particular form required by the cell and then remodel those variant forms into their native form \autocite{yi12versatility,mok13growth,keller14exogenous}. Initial binding and subsequent transport may also be sequential steps. 

\process{Predictions} For separation, receptor specificity may be influenced by the processes that cause divergence between nutrient variants. Above, I predicted that divergence caused by neutrality or functional aspects would likely lead to less separation between variants than divergence caused by escape or privacy. Less divergence between nutrient variants would lead to a greater tendency for single receptors to take up multiple variants. 

Comparative genomics and assays of receptor uptake for variants may provide opportunity to test these predictions. The patterns of amino acid substitutions in different lineages may contain information about the particular kinds of selection or neutrality that led to divergence. Those causes of divergence can be matched to the degree of receptor specificity. 

Experimental evolution studies could apply particular selective pressures and analyze the evolutionary response. For example, how do siderophores and their matching receptors evolve when the receptors change to escape viral attack?

Tradeoffs lead to evaluation of the costs and benefits for changes in design. Presumably, producing an additional uptake receptor must impose some cost. Cells benefit by adding another receptor only when the marginal gain for that new receptor exceeds the cost. Although trivial as stated, notions of cost and benefit can lead to testable hypotheses. 

For example, conditional response could lower costs by reducing expression in the absence of matching nutrients. Lower cost per receptor may associate with a greater number of receptors. Costs may be measured by competitive assays between strains with different expression attributes. In experimental evolution, higher costs may be associated with a more rapid loss of receptors in the absence of matching nutrients. 

One may also consider how high versus low nutrient availability alters the relative costs and benefits of particular receptors. Predictions about nutrient richness and receptor diversity may be tested by comparing species that live in different kinds of habitats. Additionally, experimental evolution studies can manipulate nutrient richness to test ideas about how changing marginal costs and benefits alter receptor array design.

\subsection{Species interactions}

Competitive and cooperative interactions may influence the design of receptor arrays and the flow of nutrients through communities. Interactions may be between different species or different genotypes within a species. Here, I use `species' to mean interacting types. 

\process{Specialization} A genotype can gain by taking up a broader range of variant forms. However, competing genotypes or species that specialize on a narrow range may be better at taking up particular nutrient forms. This classic generalist versus specialist tradeoff shapes the receptor arrays of individuals species.  

\process{Syntrophy} A species may use nutrients released by another species. If a common species releases a high flux of a particular nutrient form, then specialist recipients may evolve to have a narrow range of receptors.  

\process{Mutualism} A recipient may return mutualistic benefits to the genotypes that release nutrients. Such feedback may require temporal and spatial association between partners. If the nutrients passed between mutualist species become abundant, other species may hijack mutualistic flux and destroy the relation. Rarity and privacy may tend to protect a mutualism. 

\process{Predictions} Species interactions often have a game-like quality, in which the success of one species depends on what other species do. Game dynamics have many feedbacks that make it difficult to predict the full range of outcomes. Here, an outcome concerns the receptor array diversity of each species and the relative abundance of the different species. Although overall dynamics may be complex, one can sometimes make clearly defined comparative predictions.

Communities in relatively stable environments may evolve a range of specialist and generalist types \autocite{frank93evolution}. By contrast, fluctuating environments may push communities either to a diversity of specialists or to a small number of generalists. 

Species that feed syntrophically on nutrients released by common species may be more likely to have narrow, highly specialized receptor arrays.  

Mutualistic benefits between partners are more likely in spatially structured habitats. Comparing habitats, well mixed environments may have less potential for mutualisms than structured habitats. 

As mutualistic pairs become more abundant, their shared nutrients become more susceptible to hijacking by other species that do not return benefits to producers. Hijacking species may become more common with greater abundance of the mutualistic partners. Receptor array expression mediates mutualistic and hijacking relations.

\subsection{Conditional response, fluctuations and timescales} 

Cells may alter receptor expression in response to nutrient abundance and cellular need. Nutrient abundance and cellular need may fluctuate.

\process{Fluctuations} Relative timescale often determines the consequence of fluctuations. For example, loss of corrinoids probably happens by dilutive cell division and by relatively slow intrinsic decay. Thus, the timescale of need may be relatively long compared with the timescale of external fluctuations in availability. By contrast, certain species may require a relatively steady influx of glycans to fuel metabolism, setting a short timescale for need. 

\process{Sensors} To adjust receptor expression, cells must sense fluctuating nutrient availability and internal need. A sensor may respond directly to a particular nutrient. Or a sensor may respond to correlates of nutrient availability, such as time of day or rate of cellular growth and division. 

\process{Frequencies} Fluctuations and responses occur at various frequencies. For example, stochastic fluctuations of nutrient availability may happen relatively frequently, whereas daily fluxes may happen relatively infrequently. We may describe fluctuations as either a rise and fall over time or as a combination of frequencies. When using frequencies, one can think of each fluctuating aspect as a signal. 

Sensors transform external signals into internal signals. Transformation alters the frequency spectrum of the signal. Sensor design can be described in terms of tuning a filter. For example, a sensor may block high frequency stochastic fluctuations and pass through low frequency daily fluctuations---a low pass filter. The entire loop of nutrient availability, sensor information, uptake rates, and conditional response can be described in terms of frequency signal transformations. Problems of receptor array design may then be considered as an extended aspect of signal processing. 

\process{Predictions} Conditional response becomes less likely as the timescale of internal need increases relative to the timescale of external fluctuation. For example, the ratio of internal to external timescale may tend to be larger for corrinoids than for glycans, favoring greater conditional response for glycans. However, the biology of corrinoid versus glycan uptake differs in many ways, which can make such comparisons difficult to interpret. 

Comparing uptake of different glycans by the same species may provide a stronger approach. For example, the availability of one glycan may change primarily by high frequency fluctuations. By contrast, the availability of another glycan may include lower frequency fluctuations, which change over longer timescales. In general, the stronger the low frequency components of glycan fluctuation, the greater the advantage of conditional response. Thus, the response characteristics of different receptors should be tuned to the frequency spectrum of availability for the matching nutrient.

Classic control theory provides methods to optimize the response performance of a system \autocite{ogata09modern,dorf16modern}. Those methods can be used to predict attributes of uptake receptors, such as sensor design and the transformation of signals through the sequence of processes that control expression. Systems biology often uses control theory methods \autocite{alon07an-introduction,09control}. Uptake arrays provide an excellent model to test such predictions.

For example, why is expression of some glycan uptake systems stimulated by availability of the sugar components, whereas other systems only respond to the full glycan or large components of it?  The simple sugar fructose simulates expression of the uptake system for levan, a complex fructose polysaccharide. By contrast, free arabinose does not stimulate expression of the arabinan uptake systems. Only oligosaccharide components of arabinan arising from partial digestion stimulate increased expression. 

A control theory analysis of frequencies would provide insight into the correlation and mutual information between various signals. The same approach would lead to predictions about the best signals for conditional response. 

\subsection{Regulatory overwiring}

Some \textit{Bacteroides} species can take up many different glycans. Those species use a broad array of sensors to obtain information about availability. Cells integrate information from those sensors to control receptor expression. Sensor integration and control likely flow through a highly connected regulatory network. That regulatory wiring provides an excellent model to study the mechanisms by which cells integrate information and respond to their environment. 

Species differ in the variety of carbohydrates taken up. The design of regulatory wiring likely differs with diet breadth. For siderophores, cells sometimes adjust expression of uptake receptors. For corrinoids, little is known about conditional expression of uptake receptors. The different nutrient types and the different number of nutrient variants taken up by each species provide a broad basis for comparative study of regulatory wiring.

To make useful comparative predictions, we need more empirical information and conceptual understanding. On the conceptual side, I discussed analogies with deep learning. That subject focuses attention on network architecture, representation of external environmental state within the network, classification of environmental state, and control of response. 

Deep learning emphasizes the kinds of architecture and representation that allow networks to learn or to evolve by trial and error. How can we develop the vague analogies between deep learning and the regulatory wiring of cellular response? That remains an open problem. Ideally, we will develop strong comparative predictions. Here is a rough example. 

When the number of inputs and alternative environmental states is small, regulatory wiring may follow simple control loops such as basic feedback. As the number of inputs and alternative environmental states rises, regulatory wiring may become more diffuse and densely connected among a large number of nodes. In densely connected networks, it is often difficult to trace simple pathways of signal transmission and causality. Changes in environmental state cause diffuse changes in the initial layers of the network. Processing of signals through the network may cause later network layers to encode representations for components of environmental state. 

Those ideas lead to a simple prediction. When there are few variant nutrient types or environmental states, network architecture may follow simple classical control loops that can be easily parsed. As the number of variants or states increases, networks may become large and diffusely connected multilayer systems. The large networks may become \textit{overwired}, in the sense that they become more densely connected than would be designed by an engineer following classic control theory principles \autocite{frank17puzzles5}.

Many examples of basic control loops arise in simple cellular controls \autocite{alon07an-introduction,09control,somvanshi15implementation}. Larger eukaryotic regulatory networks often seem densely connected and perhaps overwired. In bacteria, it would be interesting to develop a strong comparative analysis of network architecture in relation to changes in the dimensionality and complexity of environmental challenge. For example, \textit{E.\ coli} has a relatively simple regulatory system to control response to the availability of a small number of carbohydrate food sources \autocite{kochanowski17few-regulatory}. How does the architecture of control change in species that can use an increasing number of different carbohydrate food sources?

\section{Conclusions}

I emphasized five aspects of receptor arrays. Diversification of nutrients sets the challenge for receptor uptake. Receptor specificity delimits the component properties for the array. Species interactions focus each strain on a subset of the available nutrient diversity. Fluctuations enhance the value of sensors and conditional adjustment of receptor deployment. Regulatory wiring integrates sensor information and controls the response to fluctuations.

Corrinoids, siderophores, and glycans provide opportunity for broad comparative study. Progress will enhance understanding of cellular design. Similar design problems arise in control theory and artificial intelligence. The joint study of cellular design, control, and artificial intelligence delivers synergistic benefits. 

\section*{Acknowledgments}

\noindent National Science Foundation grant DEB--1251035 supports my research.  


\begingroup
\renewcommand{\addcontentsline}[3]{}
\renewcommand{\section}[2]{}

\mybiblio	
\endgroup

\section*{Appendix}

This section lists key references for various topics. \textit{Phenotypic plasticity} analyzes an individual's adjustment of its phenotype in response to the environment. In this article, I considered how cells may adjust the expression level of different uptake receptors in response to nutrient availability. That sort of responsive phenotypic plasticity is a large subject in ecology and evolutionary biology. For example, when predators are abundant, an individual may grow protective spines \autocite{99the-ecology}. Organisms often induce various defenses in response to signals correlated with attack. Several books cover broad aspects of phenotypic plasticity and its consequences \autocite{pigliucci01phenotypic,west-eberhard03developmental,dewitt04phenotypic}.

\textit{Control theory} has played an important role in systems biology. The theory arose in engineering \autocite{davison07what,doyle09feedback,ogata09modern,dorf16modern}. Many studies of cellular regulatory control have adopted the conceptual framework of engineering control \autocite{csete02reverse,alon07an-introduction,09control,cosentino11feedback,somvanshi15implementation}. Roughly speaking, one can think of control theory in systems biology as the theoretical and mechanistic study of cellular phenotypic plasticity. 

\textit{Deep learning} comprises recent advances in computer learning systems and artificial intelligence. Those advances primarily concern techniques that greatly improve performance of computer algorithms for recognition, classification, and response to broad classes of inputs and challenges. The specific aspects of \textit{deep} concern artificial systems loosely analogous to biological neural networks. In such networks, depth expresses the number of levels of the network that connect between input nodes and output nodes. Deep networks have many levels of connectedness. The advances have to do with network architecture, representation of information, training of networks, and computer algorithms \autocite{nielsen15neural,goodfellow16deep}. For the study of regulatory control in the deployment of cellular receptor arrays, deep learning provides potentially useful analogies, insights, and computational methods.

\textit{Cellular perception} concerns cellular sensors and the use of information to regulate cellular expression and phenotype. With regard to this article, recent literature includes various analogies of cellular perception with aspects of artificial intelligence and deep learning. I believe these kinds of analogies may eventually be developed in a useful way. However, the work so far has accomplished little beyond establishing the possible relations between disciplines \autocite{lyon15the-cognitive,baluska16on-having,mitchell16cellular}.

\ifmulticol\end{multicols}\fi
\end{document}